\documentclass[12pt, appendixfloats, numberedappendix]{emulateapj}
\usepackage{amsmath}

\def\eg{e.g., }

\newcommand{\beq}{\begin{equation}}
\newcommand{\eeq}{\end{equation}}
\newcommand{\bs}[1]{\boldsymbol{#1}}
\newcommand{\norm}[1]{\left\lVert#1\right\rVert}

\usepackage{natbib}
\usepackage{rotating}
\begin{document}

\shorttitle{Nonnegative Matrix Factorization}
\shortauthors{Zhu}
\title {Nonnegative Matrix Factorization (NMF) \\
with Heteroscedastic Uncertainties and Missing data}

\author{
Guangtun Ben Zhu\altaffilmark{1,2}
} 
\altaffiltext{1}{Department of Physics \& Astronomy, Johns Hopkins University, 3400 N. Charles Street, Baltimore, MD 21218, USA, guangtun@jhu.edu}
\altaffiltext{2}{Hubble Fellow}

\begin{abstract}
Dimensionality reduction and matrix factorization techniques are important and useful machine-learning techniques in many fields.  
\textit{Nonnegative matrix factorization} (NMF) is particularly useful for spectral analysis and image processing in astronomy.
I present the vectorized update rules and an independent proof of their convergence for NMF  with heteroscedastic measurements and missing data.
I release a Python implementation of the rules and use an optical spectroscopic dataset of extragalactic sources as an example for demonstration.


\end{abstract}

\keywords{surveys -- methods: data analysis -- methods: statistical -- methods: observational}

\section {Introduction}


Astronomy has always been at the forefront of intensive data analysis.
Among the machine-learning techniques that are commonly used in the field, 
dimensionality reduction and matrix factorization techniques are particularly useful,
as astronomical observations often produce measurements with highly-correlated dimensions.
We can extract the most essential physics from the few \textit{intrinsic} dimensions revealed with the techniques. 
They also provide a powerful tool for extrapolating the data from current observations and 
making predictions for future experiments.

The most popular dimensionality reduction technique is probably principal component analysis (PCA).
Given a dataset, PCA finds the orthogonal dimensions and ranks them according to 
their contributions to the sample variance.
However, a limitation of the standard PCA is that the sample variance includes both intrinsic variance (among true values of instances)
and measurement uncertainties (for a given instance) in real-life applications.  
In astronomy, a few works have improved upon the standard PCA and developed iterative methods to 
handle the heteroscedasticity of astronomical data, usually based on the Expectation-Maximization 
optimization \citep[\eg][]{connolly1999a, budavari2009a, bailey2012a}. 

An alternative method to PCA is nonnegative matrix factorization (NMF), 
which restricts the dimensions to the nonnegative half-space.
It can be particularly useful in astronomy, since the flux of an object does not go negative (in principle). 
In their seminal paper, \citet{lee2001a} introduced simple multiplicative update rules for standard NMF with homoscedastic data.
Later, \citet{blanton2007a} presented new update rules that can take into account nonuniform uncertainties and missing data.
\citet{blondel2008a} presented a vectorized version and a proof following the same methodology as \citet{lee2001a} for homoscedastic data.
I here present an independent study of the vectorized rules and a proof for the convergence of the weighted cost function
  under these rules, by extending the original proof by \citet{lee2001a}.
I also provide a summary of the relevant formulas.
The vectorized rules are very straightforward to implement in a modern vector language. 
I here release a Python implementation of the algorithm and use an optical spectroscopic dataset of extragalactic sources 
as an example for demonstration.


\section{Nonnegative Matrix Factorization}\label{sec:nmf}

Nonnegative matrix factorization is, for a given \textit{nonnegative} matrix $\bs{X}$,
find \textit{nonnegative} matrix factors $\bs{W}$ and $\bs{H}$ such that 

\begin{equation} 
\bs{X} \approx \bs{W}\bs{H} \,\mathrm{.}
\end{equation}

\noindent We will assume the dimensions of $\bs{X}$, $\bs{W}$ and $\bs{H}$ are $l \times m$,
$l \times n$, and $n \times m$, respectively, where $n$ is a free parameter to be 
determined by the user. When discussing any given row, column, or element, we will
use $i$, $j$, and $k$ to indicate the indices in the dimensions specified by $l$, $m$, and $n$, respectively.

In the PCA language, we often interpret one of the two factors as the set of 
basis components\footnote{I here do not call NMF components eigen-components or eigen-vectors, since they are not orthogonal to each other 
because of the nonnegativity constraint.  I also do not label them principal components to avoid confusion with components from PCA, 
though I note that one can still rank the basis components by their contributions to the variance within the dataset.} 
and the other as the coefficients.
In practice, which one is which depends on how one interprets the dimensions of the original data matrix $\bs{X}$. 
In addition, a trivial transpose operation shows the symmetry between the two factors:

\begin{equation} 
\bs{X}^T \approx \bs{H}^T\bs{W}^T \,\mathrm{,} 
\label{eq:transpose}
\end{equation}

\noindent and after swapping the rows and columns all the discussions below still apply.

\subsection{The update rules for homoscedastic data}\label{sec:homorules}

To find the two factors $\bs{W}$ and $\bs{H}$, as PCA, the goal of NMF
is to minimize the least squares error (squared Frobenius norm) as the cost function:

\begin{eqnarray}
{\chi}^2 & = & \norm{(\bs{X}-\bs{W}\bs{H})}^2  \label{eq:cost} \\
 & = & \sum_{ij} \left(X_{ij} - \sum_k W_{ik}H_{kj} \right)^2 \,\mathrm{,}
\end{eqnarray}

\noindent assuming homoscedasticity of the data.

\citet{lee2001a} showed that this cost function is nonincreasing under the following multiplicative
update rules: 

%

\begin{equation}
\bs{H} \leftarrow \bs{H} \circ \frac{\bs{W}^T\bs{X}}{\bs{W}^T\bs{W}\bs{H}} \,\mathrm{,} \label{eq:Hupdate}\\
\end{equation}

\begin{equation} 
\bs{W} \leftarrow \bs{W} \circ \frac{\bs{X}\bs{H}^T}{\bs{W}\bs{H}\bs{H}^T} \,\mathrm{.} \label{eq:Wupdate}
\end{equation}

\noindent 
\noindent where $\circ$ represents the element-wise product (the Hadamard product),\footnote{The Hadamard product is commutative, 
associative and distributive over addition.} and $\frac{\,(\ )\,}{\,(\ )\,}$ the element-wise division.

It is worth pointing out that the above rules for $\bs{H}$ and $\bs{W}$ are independent of each other,
and under each of them (with the other one fixed) the cost function $\chi^2$ is guaranteed to be nonincreasing, 
as shown in their seminal paper by \citet{lee2001a}.
This means if we are trying to learn the two factors simultaneously, we need to apply them in sequential order, 
i.e., once we update $\bs{H}$, we need to use the new updated $\bs{H}$ when updating $\bs{W}$ in the next step, and vice versa, 
although which one goes first does not matter.\footnote{We can update $\bs{H}$ ($\bs{W}$) many times before updating $\bs{W}$ ($\bs{H}$).}
As another consequence, a great advantage of these rules is that, if we have known one of the two factors and are only interested in learning
the other, we can just use the corresponding rule as a projection mode while simply ignoring the other.

As many other optimization methods, the update rules above can converge to a local minimum and are not guaranteed to yield the optimal solution.
In practice, we can generate many different initializations and select the best solution. 
On the other hand, we can also increase the free parameter $n$, which is often interpreted
as the number of components, and find a better solution that reduces the cost.

\subsection{The update rules for heteroscedastic data}\label{sec:rules}

\subsubsection{The update rules}\label{sec:heterorules}

The cost function and update rules above assume homoscedastic data. 
In astronomy, we often face the challenge with nonuniform uncertainties and missing data.
To account for the heteroscedasticity, we therefore should minimize the weighted cost function:

\begin{eqnarray}
{\chi}^2 & = & \norm{\bs{V}^{\frac{1}{2}} \circ \left(\bs{X}-\bs{W}\bs{H}\right)}^2  \label{eq:weightedcost} \\
 & = & \sum_{ij} \left(V_{ij}^{\frac{1}{2}}X_{ij} - V_{ij}^{\frac{1}{2}} \sum_k W_{ik}H_{kj} \right)^2 \,\mathrm{,}
\end{eqnarray}

\noindent where the weight $\bs{V}$ has the same dimension as $\bs{X}$ and a common choice is the inverse variance matrix ${1}/{\bs{\sigma}^2}$,
and the square and square root apply element-wise. If we have missing data, we can simply use a binary mask matrix $\bs{M}$,
in which a 0-valued (\texttt{False}) element indicates a missing datum, and use $\bs{V} \circ \bs{M}$ as the new weight.

To minimize the new weighted cost function, \citet{blanton2007a} presented new update rules 
(see A22 and A24 in their appendix), which uses indexed elements explicitly.
\citet{blondel2008a} presented a vectorized version of the update rules, and also a proof following the same methodology as \citet{lee2001a}.
I here present an independent study of the vectorized rules and also provide an alternative proof in the next section, built upon the one by \citet{lee2001a}.

The vectorized update rules are
\begin{equation}
\bs{H} \leftarrow \bs{H} \circ \frac{\bs{W}^T\left(\bs{V} \circ \bs{X}\right)}{\bs{W}^T\left[\bs{V} \circ \left(\bs{W}\bs{H}\right)\right]} \,\mathrm{,} 
\label{eq:Hupdate_weighted}
\end{equation}

\begin{equation} 
\bs{W} \leftarrow \bs{W} \circ \frac{\left(\bs{V} \circ \bs{X}\right) \bs{H}^T}{\left[\bs{V} \circ \left(\bs{W}\bs{H}\right)\right]\bs{H}^T} \,\mathrm{.} 
\label{eq:Wupdate_weighted}
\end{equation}

\noindent Compared to the original rules (Eq.~\ref{eq:Hupdate} and Eq.~\ref{eq:Wupdate}), 
we have simply replaced $\bs{X}$ (in the numerators) and $\bs{W}\bs{H}$ (in the denominators) 
with the weighted versions, $\bs{V} \circ \bs{X}$ and $\bs{V} \circ (\bs{W}\bs{H})$.
I provide a proof below that under each of these rules the new weighted cost function is nonincreasing.

\subsubsection{A Proof}\label{sec:proof}

I here provide a proof that under the update rules Eq.~\ref{eq:Hupdate_weighted} and Eq.~\ref{eq:Wupdate_weighted},
the weighted cost function Eq.~\ref{eq:weightedcost} is nonincreasing, assuming the theorems in \citet{lee2001a} hold.
I refer the reader to \citet{blondel2008a} for a different proof, which includes proving the similar theorems as in \citet{lee2001a}. 

We first take a closer look at the original cost function Eq.~\ref{eq:cost} and the update rules Eq.~\ref{eq:Hupdate} and Eq.~\ref{eq:Wupdate}.
First I note that the update rule Eq.~\ref{eq:Hupdate} applies to each column without interference (between the columns),
i.e., the update rule is equivalent to updating every column $j$ in $\bs{H}$ 
in parallel:\footnote{This feature can therefore be used to improve the efficiency of NMF with parallel programming.}

\begin{equation}
\bs{H}_j  \leftarrow  \bs{H}_j \circ \frac{\bs{W}^T\bs{X}_j}{\bs{W}^T\bs{W}\bs{H}_j} \,\mathrm{,} \label{eq:Hupdate_j}
\end{equation}

\noindent and the cost function can be written as a sum of contributions from all the columns:

\begin{eqnarray}
{\chi}^2 & = & \sum_{j} \norm{(\bs{X}_j-\bs{W}\bs{H}_j)}^2  \\
         & = & \sum_{j} {\chi}^2_j \,\mathrm{.}
\end{eqnarray}

\noindent 
This is the strategy used by \citet{lee2001a}, who provided the proof for the theorem that, the contribution 
to the cost function from a given column $j$, $\chi^2_j$, is nonincreasing under the update rule above (Eq.~\ref{eq:Hupdate_j}), 
and therefore the sum of the contributions from all the columns, $\chi^2$, is also nonincreasing. 
I refer the reader to their paper for its proof and here I will take it as a theorem, 
and use it to prove that under the following update rule for a given column $j$,

\begin{equation}
\bs{H}_j  \leftarrow  \bs{H}_j \circ \frac{\bs{W}^T\left(\bs{V}_j \circ \bs{X}_j\right)}{\bs{W}^T\left[\bs{V}_j \circ \left(\bs{W}\bs{H}_j\right)\right]}\,\mathrm{,}
\label{eq:Hupdate_weighted_j}
\end{equation}

\noindent its contribution to the weighted cost function, 

\begin{equation}
{\chi}^2_j  =  \norm{\bs{V}^{\frac{1}{2}}_j \circ (\bs{X}_j-\bs{W}\bs{H}_j)}^2 \,\mathrm{,} \\
\label{eq:weightedcost_j}
\end{equation}
\noindent is nonincreasing.

To prove this, note that the update rules for $\bs{H}$ and $\bs{W}$ are independent of each other 
and are applied with the other one fixed. When updating $\bs{H}_j$, we can define $\bs{\hat{X}}$ and
$\bs{\hat{W}}$:
\begin{eqnarray}
\bs{\hat{X}}_j & \equiv & \bs{V}^{\frac{1}{2}}_j \circ \bs{X}_j \,\mathrm{,}\\
\bs{\hat{W}} & \equiv & [\bs{V}^{\frac{1}{2}}_j]^{\mathrm{diag}}\,\bs{W} \,\mathrm{,}
\end{eqnarray}
\noindent where $[\bs{V}^{\frac{1}{2}}_j]^{\mathrm{diag}}$ represents a diagonal matrix with $\bs{V}^{\frac{1}{2}}_j$
as the diagonal elements.
%

We can now re-write Eq.~\ref{eq:weightedcost_j} as
\begin{equation}
{\chi}^2_j  = \norm{(\bs{\hat{X}}_j-\bs{\hat{W}}\bs{H_j})}^2 \,\mathrm{.}
\label{eq:cost_hat}
\end{equation}

\noindent And a little algebra shows that the update rule Eq.~\ref{eq:Hupdate_weighted_j} can be re-written as
\begin{equation}
\bs{H}_j  \leftarrow \bs{H}_j \circ \frac{\bs{\hat{W}}^T\bs{\hat{X}}_j}{\bs{\hat{W}}^T\bs{\hat{W}}\bs{H}_j} \,\mathrm{.}
\label{eq:Hupdate_hat}
\end{equation}

\noindent Now we can apply the original proof by \citet{lee2001a} and show that, 
under the rule Eq.~\ref{eq:Hupdate_hat}, the cost Eq.~\ref{eq:cost_hat} is nonincreasing.

\vspace{0.1in}
With a simple transpose operation as in Eq.~\ref{eq:transpose}, we can also show that under the update rule Eq.~\ref{eq:Wupdate_weighted}, 
the weighted cost function Eq.~\ref{eq:weightedcost} is nonincreasing. For completeness, alternatively, 
we can also decompose the cost function into contributions from every row in $\bs{X}$:
\begin{eqnarray}
{\chi}^2 & = & \sum_{i} \norm{\bs{V}^{\frac{1}{2}}_i \circ (\bs{X}_i-\bs{W}_i\bs{H})}^2  \\
         & = & \sum_{i} {\chi}^2_i \,\mathrm{.}
\end{eqnarray}

\noindent By defining
\begin{eqnarray}
\bs{\hat{X}}_i & \equiv & \bs{V}^{\frac{1}{2}}_i \circ \bs{X}_i \,\mathrm{,}\\
\bs{\hat{H}} & \equiv & \bs{H}\,[\bs{V}^{\frac{1}{2}}_i]^{\mathrm{diag}} \,\mathrm{,}
\end{eqnarray}
\noindent where $[\bs{V}^{\frac{1}{2}}_i]^{\mathrm{diag}}$ represents a diagonal matrix with $\bs{V}^{\frac{1}{2}}_i$ as the diagonal elements.
we can re-write the update rule Eq.~\ref{eq:Wupdate_weighted} for a given row $i$ as
\begin{eqnarray} 
\bs{W}_i & \leftarrow & \bs{W}_i \circ \frac{\left(\bs{V}_i \circ \bs{X}_i\right) \bs{H}^T}{\left[\bs{V}_i \circ \left(\bs{W}_i\bs{H}\right)\right]\bs{H}^T} \\
       & = & \bs{W}_i \circ \frac{\bs{\hat{X}}_i\bs{\hat{H}}^T}{\bs{W}_i\bs{\hat{H}}\bs{\hat{H}}^T} \,\mathrm{,}
\end{eqnarray}

\noindent under which the contribution to the weighted cost function from row $i$,

\begin{equation}
{\chi}^2_i =  \norm{(\bs{\hat{X}}_i-\bs{W}_i\bs{\hat{H}})}^2 \,\mathrm{,}
\end{equation}

\noindent is nonincreasing.

\begin{figure}
\epsscale{1.20}
\plotone{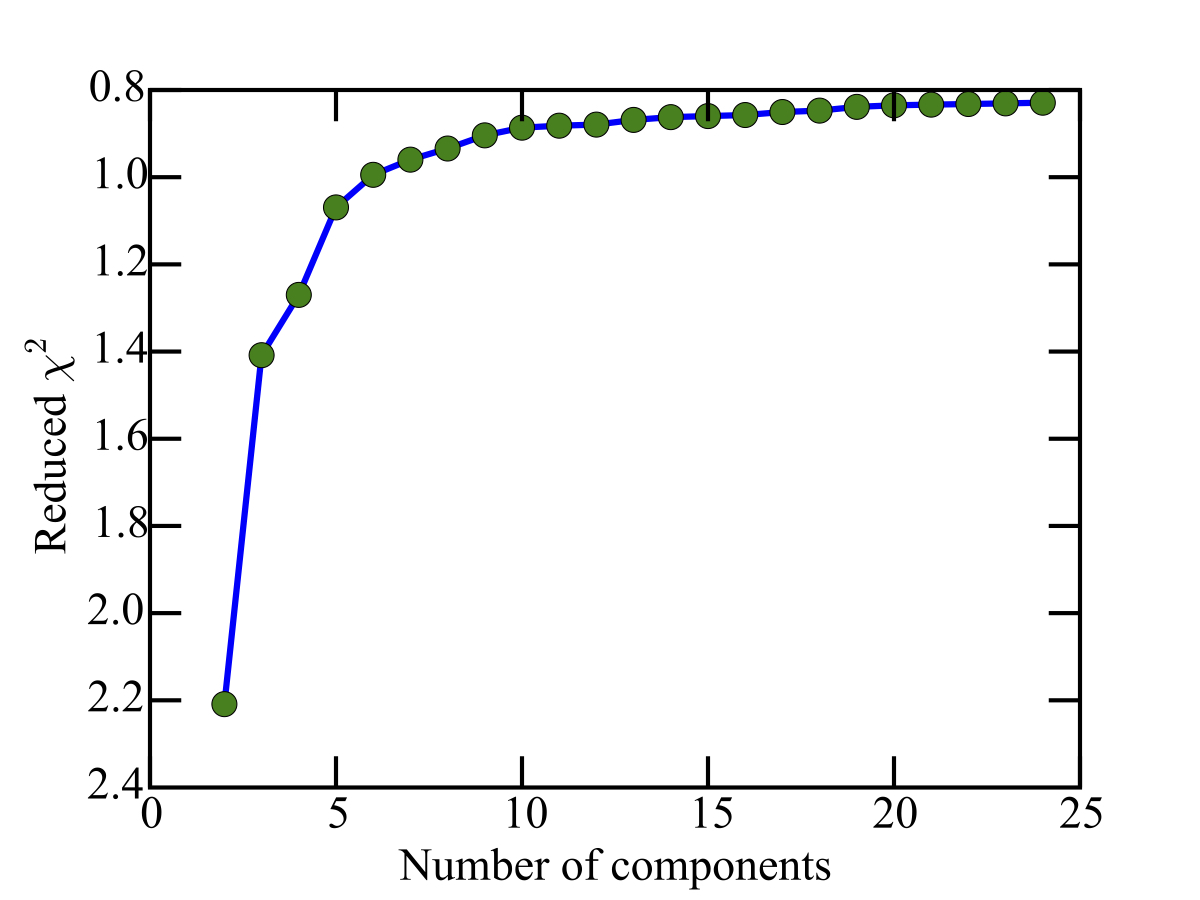}
\caption{The progression of the reduced $\chi_{\rm red}^2$ for the extragalactic spectroscopic dataset, as a function of the free parameter $n$, the number of components. 
I have divided the total weighted cost function by the data size minus the number of basis components.}
\vspace{0.2cm}
\label{fig:chi2}
\end{figure}

\begin{figure*}
\epsscale{1.05}
\plotone{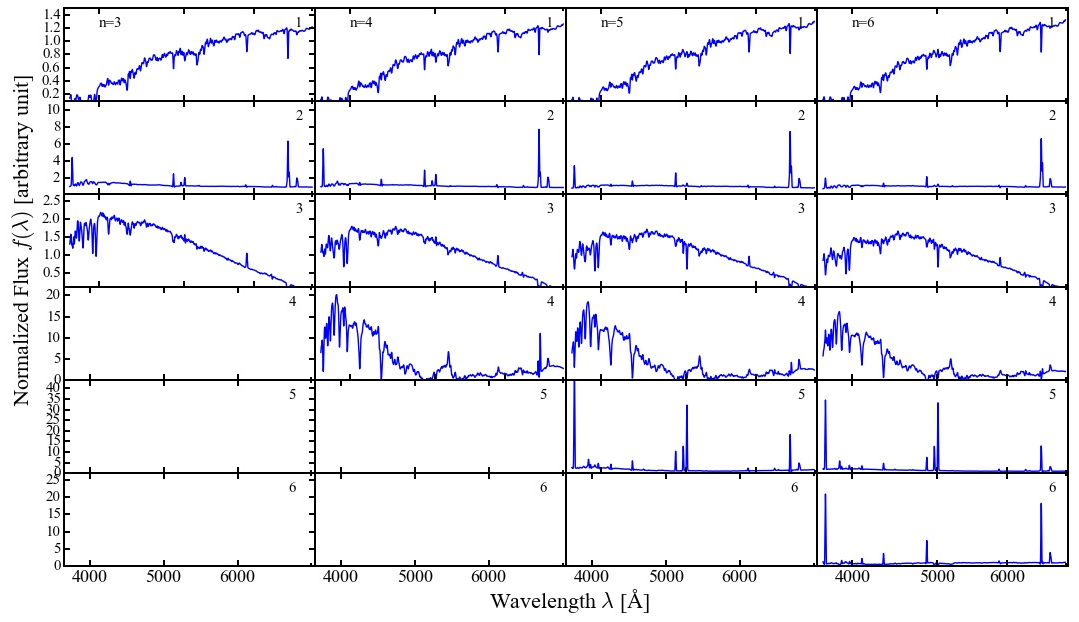}
\plotone{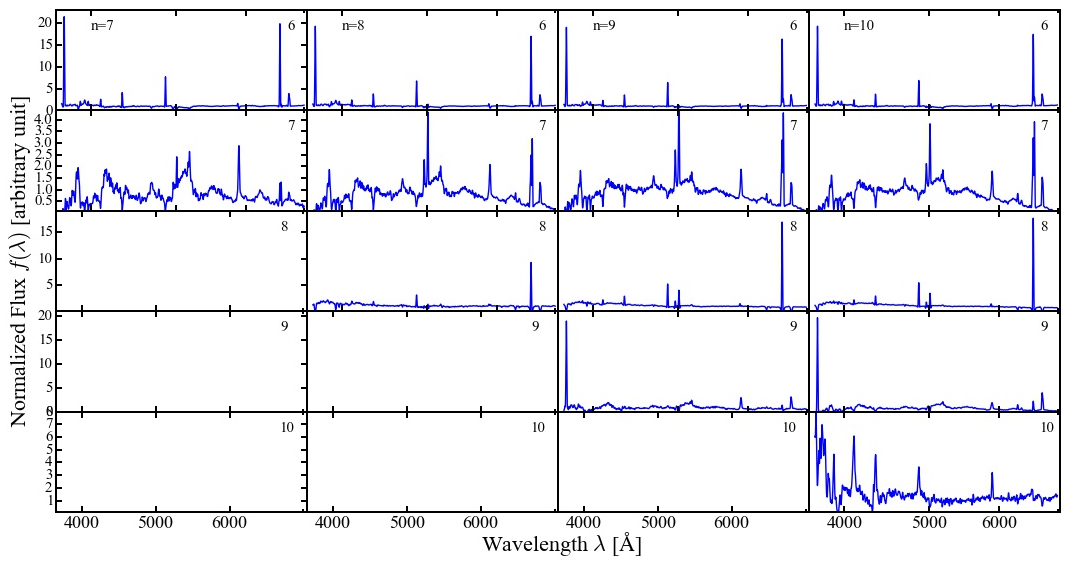}
\caption{The basis sets for the extragalactic spectroscopic test dataset, for $n=3$ to $n=10$. Note I have skipped the first $5$ components for $n=7$ to $n=10$, 
which are similar to the basis set for $n=5$.}
\vspace{0.2cm}
\label{fig:basisset}
\end{figure*}

\begin{figure*}
\epsscale{1.05}
\plotone{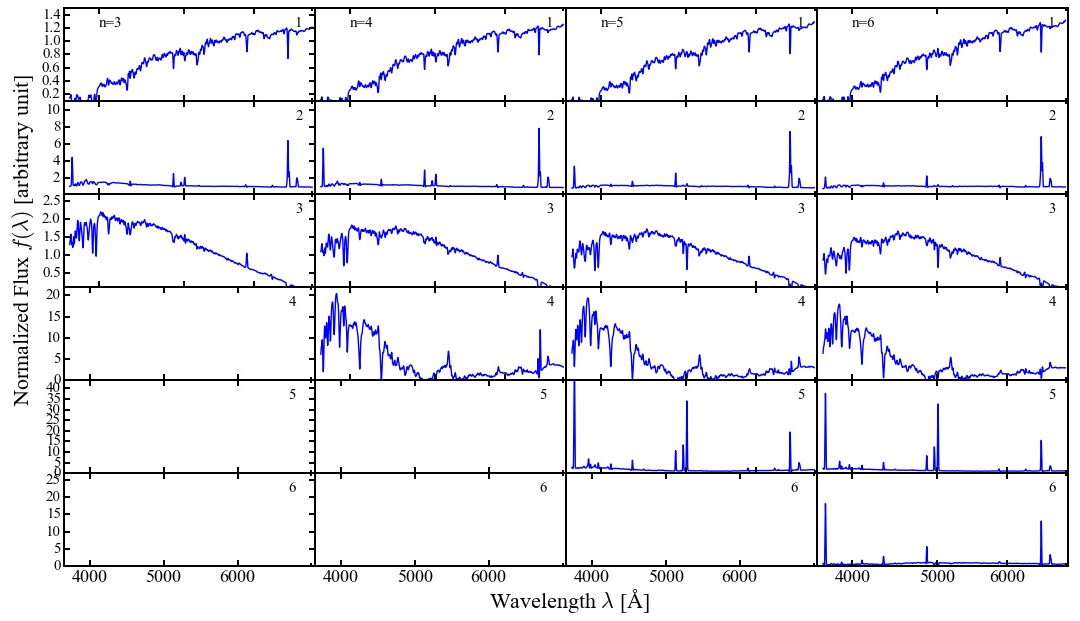}
\plotone{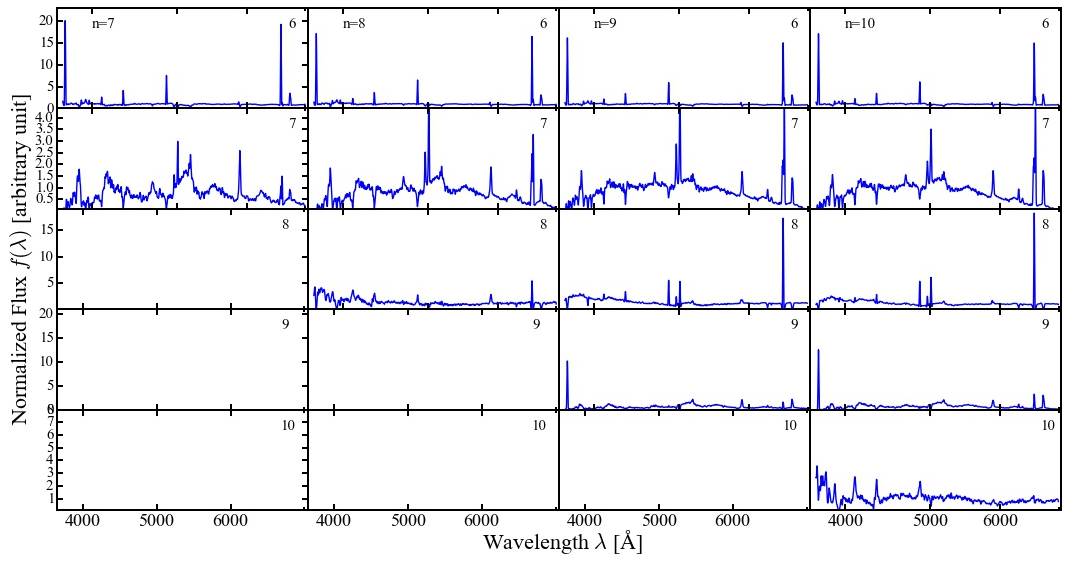}
\caption{The basis sets for the extragalactic spectroscopic test dataset, but with $20\%$ data masked, for $n=3$ to $n=10$. Note I have skipped the first $5$ components for $n=7$ to $n=10$, 
which are similar to the basis set for $n=5$.}
\vspace{0.2cm}
\label{fig:basissetmask}
\end{figure*}

\section{An example}\label{sec:example}

It is easy to implement the vectorized algorithm above in any vector language.
I here release a Python implementation, taking advantage of the fast NumPy package that has BLAS and LAPACK incorporated.
I present a brief introduction of the code in the Appendix.
Here I use an optical spectroscopic dataset of extragalactic sources as an example to demonstrate 
how to apply the technique in practice.

\subsection{The test dataset}
To further illustrate the algorithm,
I use an optical spectroscopic dataset of extragalactic sources as an example.
This dataset was also used as a test case in Zhu (2016) for classification purposes 
and can be downloaded on the Web, as shown in the Appendix.
I select the sources and their spectra from the seventh 
data release \citep[DR7,][]{abazajian2009a} of the Sloan Digital Sky Survey \citep{york2000a}.
More specifically, I select sources at redshift $z\sim0.05$ with high-S/N ($15-30$) spectra. 
The test dataset includes $2820$ extragalactic sources and 
I consider the wavelength range between $3700\,$\AA\ and $7000\,$\AA.
Although a small sample, this dataset spans a variety of astrophysical sources, 
including quiescent, star-forming galaxies with different metallicities, AGN, etc.,
and serves well for the testing purposes.

\begin{figure}
\vspace{-0.2cm}
\epsscale{1.15}
\plotone{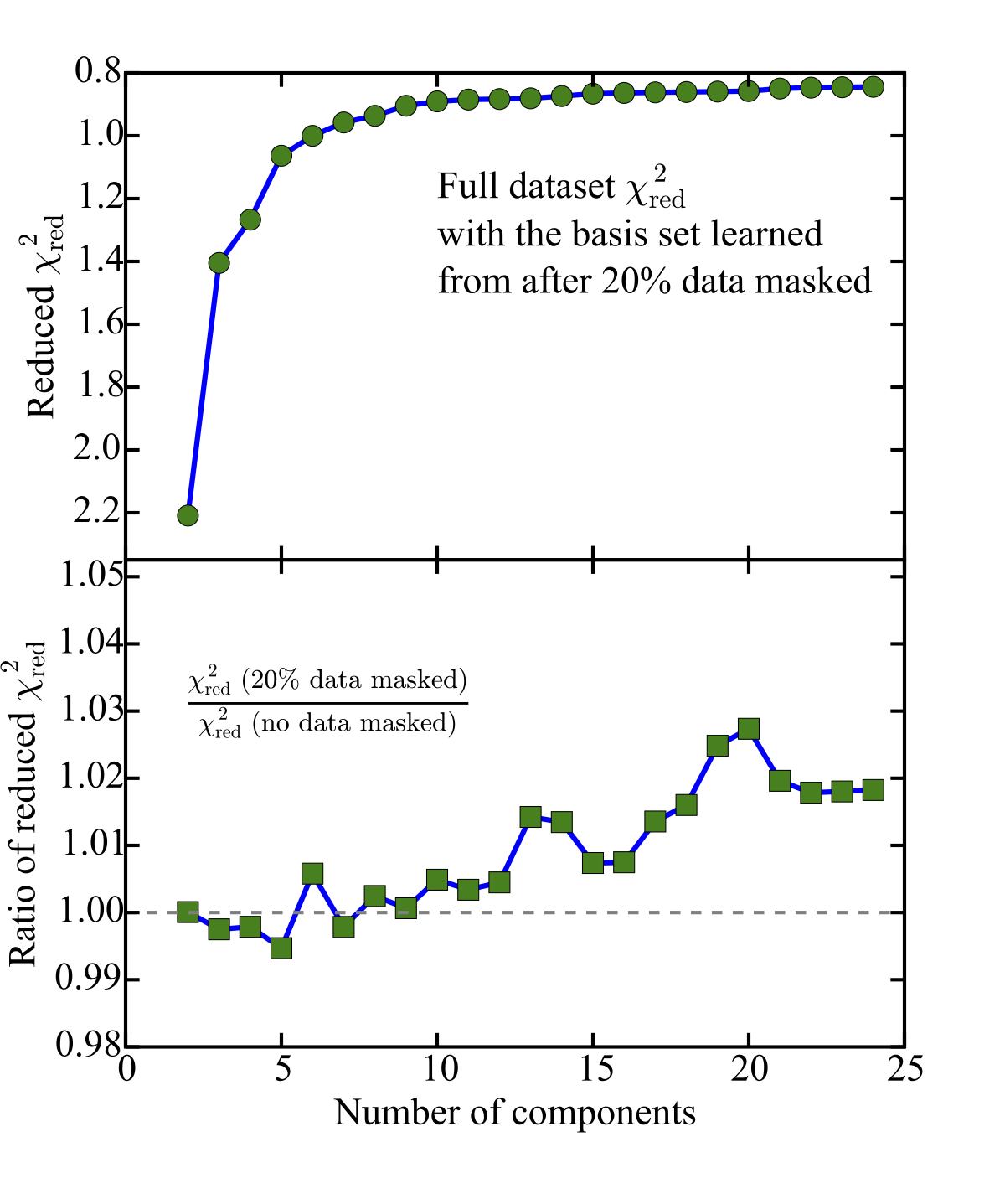}
\caption{\textit{Upper panel}: The progression of the reduced $\chi_{\rm red}^2$ for the extragalactic spectroscopic dataset, 
as a function of the free parameter $n$, the number of components.  
Here I applied to the full dataset the basis sets learned after randomly masking out $20\%$ of the data.
\textit{Lower Panel}: The ratio between $\chi_{\rm red}^2$ using the basis sets after randomly masking out $20\%$ 
of the data and that using the basis sets learned with the full dataset.}
\vspace{0.2cm}
\label{fig:chi2comp}
\end{figure}

\vspace{0.1in}

\subsection{The basis sets}
I construct basis sets with the number of components $n$ ranging from $2$ to $24$. 
For comparison purposes, for a given $n$,
I initialize the first $n-1$ components with the previous basis set, 
except for the first basis set with $n=2$, for which I generate random number matrices with values between 0 and 1 as initial guesses.
This way we create a roughly ranked list of basis components based on their contributions to the sample variance. 
\textit{However, I would like to stress that this construction method is devised only for the convenience 
of comparing the basis sets with different numbers of components}. 
As discussed earlier, the update rules of NMF are not guaranteed to converge to the global minimum of the cost function.

In Figure~\ref{fig:chi2}, I show the progression of the reduced $\chi_{\rm red}^2$, 
the total cost divided by the data size minus the number of components $n$, as a function of $n$.
The cost converges quickly and at $n=10$, it is already less than $5\%$ higher than at $n=24$.

To take a closer look, I show the basis sets for $n=3$ to $n=10$ in Figure~\ref{fig:basisset}.
Note I use the basis set for $n-1$ as the initial set at the construction 
for $n$, so the first $n-1$ components of each basis set are almost the same as the
previous basis set.  For $n=7$ to $n=10$, I have skipped the first $5$ components, 
which are very similar to the basis set for $n=5$. 
Looking at the first two components (which form the basis set for $n=2$), 
we see a red template that looks like the spectrum of a quiescent galaxy dominated by old stellar populations \citep[see, \eg][]{thomas2005a, zhu2010a} 
and a blue template that appears to be the spectrum of a star-forming galaxy with strong emission lines \citep[see, \eg][]{kennicutt1992a, zhu2015a}.
The third component adds a bluer template that appears to be a combination of B, A and F types of stars. 
The fourth and the seventh components show variations of stellar absorption features, 
likely to account for the range of stellar age and metallicity covered by the sample. 

The rest of the components mostly include emission lines with varying ratios.
This is because in different types of systems, different physical mechanisms are 
responsible for exciting the atoms/ions, resulting in a variety of emission-line properties.
These strong emission lines can account for a considerable fraction of the least squares error and 
therefore require a number of basis components with varying ratios to cover the variance.

\subsection{The basis set with partial data}

We can handle missing data easily with the algorithm described above, 
simply by providing a mask $\bs{M}$.
It is interesting to explore the effects of missing data on the modeling.
To do so, I randomly select $20\%$ of the data\footnote{I do not select random objects and exclude 
their full spectra, but instead I select random elements from the dataset as a whole.} 
and assume they are corrupted.

Applying the same procedure as above for the full dataset,
I construct basis sets as a function of the number of components.
Figure~\ref{fig:basissetmask} presents the basis sets for $n=3$ to $n=10$.
Remarkably, they are very similar to those basis sets learned from the full dataset. 
A careful comparison shows that there are some small differences on a few percent level,
but it is reassuring that the algorithm can achieve such robust results.

To take a further look, I take the basis sets built from the partial dataset,
and apply them to the full dataset to see how well they can describe the full dataset as a whole.
The projection mode (with one factor fixed) of the algorithm makes this a trivial task.
In the upper panel of Figure~\ref{fig:chi2comp}, I show the progression of the reduced $\chi_{\rm red}^2$ from this experiment. 
In the lower panel, I compare the cost to that with the basis sets learned from the \textit{full} sample.
The result shows that the basis sets built from the partial dataset, with $20\%$ data masked, 
can account for the variance in the full dataset up to $98\%$. 

Because the algorithm can handle missing data, it can be very useful in extragalactic astronomy.
When we observe sources at different redshifts with a given instrument, 
the fixed observer-frame wavelength coverage translates to a running rest-frame window.
If we use standard PCA or the original NMF, then we need to restrict the analysis to a 
small rest-frame wavelength range in which sources at different redshifts have common coverage.
With the NMF technique described here, \citet{zhu2013a} modeled quasar spectra over a large redshift 
and rest-frame wavelength range (e.g., see their Figure~14), and I refer interested readers 
to that paper for more information about the application.

\section{Summary}\label{sec:summary}

With the amount of data growing exponentially, one of the major challenges 
in modern astronomy is how to efficiently extract scientific information.
I presented a simple vectorized algorithm for nonnegative matrix factorization (NMF) 
with nonuniform uncertainties and missing data, which can be easily implemented in any modern vector language. 
I released a Python implementation of the algorithm. 
Using an optical spectroscopic dataset of extragalactic sources as an example, 
I showed how the weighted cost progresses and converges as a function of the number of components, 
the only free parameter in the technique. I have also discussed how to take advantage of 
the independent, sequential update rules for the two factors.

As a final note, although techniques such as PCA and NMF are powerful tools that 
can reduce the dimensionality of the data and reveal intrinsic dimensions informative 
about the underlying correlations and physics, they still have some fundamental limitations.
For example, the (linear) combinations of the basis components 
can occupy a part of the space that is otherwise empty in reality.
As another example, these techniques attempt to find the basis components that dominate the variance (so as to minimize the least squares error), 
and a small number of extreme cases are often not included in the modeling.
In astronomy, these cases are often the most interesting ones as they present the best 
opportunities for groundbreaking discoveries.
These limitations could be catastrophic especially when we are facing low S/N data,
which has become more and more common in the new era featuring large surveys.
To overcome these limitations, we need to resort to other techniques, 
such as using a set of archetypes to represent all the instances in the dataset \citep[\eg][]{zhu2016a}.

\acknowledgments

\vspace{0.3in}
I acknowledge support provided by NASA through Hubble Fellowship grant \#HST-HF2-51351 awarded by the Space Telescope Science Institute, 
which is operated by the Association of Universities for Research in Astronomy, Inc., under contract NAS 5-26555. 

This article uses the public data from the SDSS survey. Funding for the SDSS and SDSS-II has been provided by the Alfred P. Sloan Foundation, the Participating Institutions, the National Science Foundation, the U.S. Department of Energy, the National Aeronautics and Space Administration, the Japanese Monbukagakusho, the Max Planck Society, and the Higher Education Funding Council for England. The SDSS Web Site is http://www.sdss.org/.

The SDSS is managed by the Astrophysical Research Consortium for the Participating Institutions. The Participating Institutions are the American Museum of Natural History, Astrophysical Institute Potsdam, University of Basel, University of Cambridge, Case Western Reserve University, University of Chicago, Drexel University, Fermilab, the Institute for Advanced Study, the Japan Participation Group, Johns Hopkins University, the Joint Institute for Nuclear Astrophysics, the Kavli Institute for Particle Astrophysics and Cosmology, the Korean Scientist Group, the Chinese Academy of Sciences (LAMOST), Los Alamos National Laboratory, the Max-Planck-Institute for Astronomy (MPIA), the Max-Planck-Institute for Astrophysics (MPA), New Mexico State University, Ohio State University, University of Pittsburgh, University of Portsmouth, Princeton University, the United States Naval Observatory, and the University of Washington.

\appendix

\section{The code}\label{app:code}

\subsection{Install the package}

I have implemented the algorithm for NMF described in Section~\ref{sec:heterorules} in Python 3, 
including options for sparse matrices and  projection mode.
I share the code, with the name \texttt{NonnegMFPy}, on the repository hosting service GitHub.
Interested user can fork or clone the repository.\footnote{\texttt{https://github.com/guangtunbenzhu/NonnegMFPy}}
For readers who are only interested in using the package, I have also released the code on Python Package Index (PyPI)
and one can install it with:

\begin{verbatim}
  > pip install NonnegMFPy
\end{verbatim}

We recommend installing it via pip unless the user is keen in helping further develop and maintain the package.
The repository webpage also includes documentation and a user guide of the package. 
The user can provide weight ($\bs{V}$) and mask ($\bs{M}$) matrices to take into account the uncertainties and missing data.
In addition, the code also includes projection mode options, \texttt{H\_only} and \texttt{W\_only}, to learn only $\bs{H}$ or $\bs{W}$,
given $\bs{W}$ or $\bs{H}$.

\subsection{Test the code}

The test optical spectroscopic dataset I used in the paper is publicly available\footnote{See README on \texttt{https://github.com/guangtunbenzhu/NonnegMFPy}}
in \texttt{fits} format\footnote{\texttt{http://fits.gsfc.nasa.gov/}}. 
Assuming we have extracted the spectra matrix \texttt{spec} and the inverse variance matrix \texttt{ivar} from the dataset.
The user can test the code as follows.

\begin{verbatim}
 > from NonnegMFPy import nmf
 > g = nmf.NMF(spec, V=ivar, n_components=5)
 > chi2, time_used = g.SolveNMF()
\end{verbatim}
 
If the user has already learned $\bs{W}$ and wishes to learn $\bs{H}$ only, one can simply set \texttt{H\_only = True}:
\begin{verbatim}
 > g = nmf.NMF(spec, V=ivar, W=W_known, n_components=5)
 > chi2, time_used = g.SolveNMF(H_only=True)
\end{verbatim}
Note the dimensions need to be compatible.

\bibliographystyle{apj}


\end{document}